\documentclass[showpacs,preprintnumbers,amsmath,amssymb,12pt,floatfix,epsfig]{revtex4}
\usepackage{amssymb}
\usepackage{bm}
\usepackage{graphicx}
\def\be{\begin{equation}}
\def\ee{\end{equation}}
\def\Be{\begin{eqnarray}}
\def\Ee{\end{eqnarray}}
\def\ba{\begin{array}}
\def\ea{\end{array}}

\begin{document}
\draft

\title{Roles of High-lying Excited States on Neutrino Reactions and the Gamow Teller strength for $^{40}$Ar}
\author{ Eunja Ha and Myung-Ki Cheoun \footnote{Corresponding author : cheoun@ssu.ac.kr}}
\address{ Department of Physics, Soongsil University, Seoul 156-743, Korea }


\begin{abstract}
Neutrino reactions on $^{40}$Ar via charged and neutral
currents for detecting solar and core
collapsing supernovae (SNe) neutrinos and the Gamow Teller strength are calculated by considering the
high-lying excited states up to a few tens of MeV region. The nucleus
was originally exploited to identify the solar neutrino emitted
from $^{8}$B produced in the pp-chains on the Sun. With the higher
energy neutrinos emitted from the core collapsing SNe,
contributions from higher multi-pole transitions including the
spin dipole resonances (SDR) as well as the Gamow Teller (GT) and Fermi
transitions are shown to be important ingredients for
understanding reactions induced by the SNe as well as solar neutrinos. In this work,
we focused on the role of high-lying excited states which are located
beyond a few low-lying states known in the experiment. Expected large
difference between the cross sections of $\nu_e$ and ${\bar
\nu_e}$ reactions on $^{40}$Ar, which difference has been anticipated in previous calculations
because of the large Q value in the ${\bar \nu}_e$ reaction, is
significantly diminished. The reduction leads only to about 2
times difference between them. Our calculations are carried out by
the Quasi-particle Random Phase Approximation (QRPA), which takes
account of the neutron-proton pairing as well as proton-proton and
neutron-neutron pairing correlations. They were successfully
applied in the description of the nuclear beta decay and relevant neutrino
reaction data on $^{12}$C and $^{56}$Fe, and the GT data on the $^{138}$La and
$^{180}$Ta.
\end{abstract}

\date{\today}
\maketitle

\section{Introduction}
Neutrino reactions on $^{40}$Ar were of astrophysical importance
because they may be used to detect the solar neutrino
emitted from $^{8}$B via the pp-chains in the Sun. The reactions are measured through the
liquid argon time projection chamber (LArTPC) at the ICARUS
(Imaging of Cosmic and Rare Underground Signals) \cite{Rub98}.
Since the maximum energy of the solar neutrino is thought to be
about 17 MeV in the standard solar model, the neutrino reactions
are naturally sensitive on the low-lying discrete energy states of $^{40}$Ar.

Since the protons in $^{40}$Ar are occupied mostly at the $sd$ shell while the neutrons are located up to the $pf$ shell,
charge exchange reactions (CEX) on the nucleus are closely related to the multi-particle and multi-hole interactions of the both major shells.
For instance, the Q-value
for the $^{40}$Ar($\nu_e, e^-)$$^{40}$K$^{*}$ reaction is 1.50
MeV, while the Q-value for the $^{40}$Ar(${\bar \nu}_e, e^+)$$^{40}$Cl$^{*}$
reaction is 7.48 MeV, 5 times larger than the $\nu_e$ reaction. Therefore, $^{40}$Ar(${\bar \nu}_e,
e^+)$$^{40}$Cl$^{*}$ reactions might be kinematically disfavored
in the low energy neutrino, such as solar neutrino, reactions. Moreover, in $^{40}$Cl$^{*}$, only 2 excited states for the Gamow
Teller (GT) transition are known with no excited isobaric analogue
states (ISB). In this
respect, $^{40}$Ar was thought to effectively distinguish the
$\nu_e$ and ${\bar \nu}_e$ emitted from the Sun.

Recently, authors in Ref. \cite{Rub08} revised their
previous works \cite{Rub03} in order to discuss the possible detection
of neutrino oscillation of supernova (SN) neutrinos. The neutrinos
emitted from the SN explosion can give valuable information about neutrino properties, such as the $\nu$ mixing angle $\theta_{13}$
and the mass hierarchy, because they traverse regions of dense
matter in the exploding star where matter enhanced oscillations
take place.

Here we briefly discuss
such feasibility about how to extract such neutrino properties. One possible way is to investigate the
abundances of light nuclei. Among the light nuclei, $^{7}$Li and
$^{11}$B are abundantly produced through the $\nu$-process ($\nu$-induced reactions on related nuclei in the core
collapsing supernova) \cite{Woosley90,Yoshida08,Heg,Suzuki09}.
Since the $\nu$-induced reaction might be sensitive on the
$\nu$-properties as well as $\nu$-flavors, the $^{7}$Li and
$^{11}$B abundances could be sensitive on the $\nu$ parameters. For
example, in Refs. \cite{Yoshida08,Suzuki06}, the production yields
of $^{7}$Li and $^{11}$B are shown to be sensitive on the neutrino mass
hierarchy as well as the emitted neutrino temperature, although some other interpretations are still remained \cite{Ch11}.

Another way is to directly detect the $\nu$ signals coming from
the core collapsing SN explosion on the Earth \cite{Smir00}. The
LArTPC was suggested as one of possibilities to enable such
detection. This ICARUS-like detector is also planned to detect the
neutrino beam at CERN \cite{Curi06}. Since neutrino energies from
the SN explosion are expected to have tens of MeV, which are higher than those stemmed from
the solar neutrino \cite{Woosley90,Yoshida08}, one needs to
consider the contributions from higher multi-pole transitions as well as high-lying excited states.

Refs. \cite{Rub08,Rub03} exploited the RPA calculation done by
Kolbe and Langanke \cite{Kolbe03-a} and showed that contributions
from higher multipoles, such as spin dipole resonances (SDR),
could be important for the SN neutrino. The importance of higher
multipole transitions is recently confirmed at the shell model
(SM) calculation by Suzuki {\it et al.} \cite{Suzuki06} and also
at the Quasi-particle Random Phase Approximation (QRPA)
calculations by us \cite{Ch09-1,Ch09-2,Ch10}.

Neutrino ($\nu$) (antineutrino (${\bar \nu}$)) energies and flux
emitted from the core collapsing SN explosion are conjectured to be
peaked from a few to tens of MeV energy region
\cite{Woosley90,Yoshida08,Burrow}. Therefore, the $\nu ({\bar
\nu})$-induced reactions on $^{40}$Ar are sensitive on the high-lying
excited states of the nucleus beyond one nucleon thresholds. One more point to be noticed is that the reaction
proceeds via two-step processes, {\it i.e.} target nuclei are
excited by incident $\nu ({\bar \nu}) $ and decayed to lower
energy states with the emission of some particles
\cite{Kolbe03-a}. Of course, the one-step process, which directly knocks out one nucleon from a target nucleus with outgoing lepton, might work on the abundances of final nuclei \cite{Ch09-3}. In particular, the contribution could be comparable to that of the two-step process with the higher energy neutrino \cite{Ch09-4}.  But, in this work, we do not take account of the effects by the one-step process.

The particle emission decay mode allows the daughter nuclei to
easily proceed to adjacent nuclei and naturally influences the
nuclear abundances in the universe. Since we do not have
experimental data about the emission decay mode enough to fix
relevant transitions, one usually resorts to the statistical model
such as Hauser-Feshbach theory \cite{Haus52,Naka05,Suzuki06}. In this work, we take the approach done by Ref. \cite{Naka05}. Of course,
the preceding excitation occurs through various transitions, {\it
i.e.} super allowed Fermi ($J^{\pi} = 0^+)$, allowed Gamow Teller
(GT) ($J^{\pi} = 1^+)$, SDR ($J^{\pi} = 0^- , 1^-, 2^-)$, and
other higher multipole transitions \cite{Suzuki06,Ring08}.

Since out results for the SN neutrinos are briefly reported \cite{Ch11-2}, in this work, we present more detailed and advanced results based on the QRPA calculation
for $\nu ({\bar \nu})$-$^{40}$Ar reactions in the following two respects. The first point is that relevant cross
sections for solar and SN neutrinos are calculated and compared simultaneously, whose energy
ranges are considered up to 30 and 80 MeV region, respectively
\cite{Burrow}. Secondly, we focus on roles of high-lying excited
states around 20 MeV although they are not verified at the
experiments. Our results decrease the cross sections of
$^{40}$Ar($\nu_e, e^-)$$^{40}$K$^{*}$ reaction about 3.5 times and
increase about twice those of $^{40}$Ar(${\bar \nu}_e,
e^+)$$^{40}$Cl$^{*}$ reaction compared to the previous
calculations \cite{Rub08,Rub03}. Consequently, the expected 12
times difference between the reactions at $E_{\nu}$ = 80 MeV is
reduced drastically to about twice difference. The GT strength
recently deduced from $^{40}$Ar(p,n) reactions \cite{Bhatt09},
which is estimated to justify our calculations, is
also reproduced in a satisfactory way.

In section II, our theoretical formalism is presented with the
neutron-proton pairing correlations. Results of various neutrino
induced reactions on $^{40}$Ar are discussed at section III. Finally
summaries and conclusion are given at section IV.

\section{Theoretical Frameworks}

Since our QRPA formalism for neutrino-nucleus ($\nu - A$)
reactions is detailed at our previous papers \cite{Ch09-1,Ch09-2},
here we summarize two important characteristics compared to other
QRPA approaches. First, the Brueckner ${\cal G}$ matrix is
employed for two-body interactions inside nuclei by solving the
following Bethe-Salpeter equation based on the Bonn CD potential
for nucleon-nucleon interactions in free space
\begin{equation} {\cal G}(w)_{ab,cd}  = V_{ab,cd} +
 V_{ab,cd} { Q_{p} \over {w
- H_{0}}} {\cal G}(w)_{ab,cd}~, \end{equation}
where $a, b, c,d$ indicate the single nucleon basis states
characterized by oscillator type wave functions with single
particle energies from the Woods-Saxon potential. $H_{0}$ is the
harmonic oscillator Hamiltonian and $Q_{p}$ is the Pauli operator.
$V_{ab,cd}$ is the phenomenological nucleon-nucleon potential in
free space. It may enable us to reduce some ambiguities from
nucleon-nucleon interactions inside nuclei.

Second, we include neutron-proton (np) pairing as well as
neutron-neutron (nn) and proton-proton (pp) pairing correlations.
In medium or medium-heavy nuclei, the np pairing contributes to
some extent for relevant transitions because of small energy gaps
between proton and neutron energy spaces \cite{Ch93}. Moreover the
np pairing leads to a unified description of both charged current
(CC) and neutral current (NC) reactions within a framework as
shown later on.

But the contribution by the np pairing is shown to be only within
1 $\sim$ 2 \% for the weak interaction in $^{12}$C, such as
$\beta^{\pm}$ decay and the $\nu - ^{12} $C reaction
\cite{Ch09-1,Ch09-2}, because the energy gap between neutron and
proton energy spaces is too large in such a light nucleus to be
effective. But in medium-heavy nuclei, such as $^{56}$Fe and
$^{56}$Ni, the np pairing effect accounts for 20 $\sim$ 30 \% of
total cross sections \cite{Ch09-2}. Results by our QRPA have
successfully described relevant $\nu -$reaction data for $^{12}$C
\cite{Ch09-1}, $^{56}$Fe and $^{56}$Ni \cite{Ch09-2}, and
$^{138}$La and $^{180}$Ta \cite{Ch10} as well as $\beta$, 2$\nu
\beta \beta$ and $0 \nu 2 \beta$ decays \cite{Ch93}.

For CC reactions, the ground state of a target nucleus is
described by the BCS vacua for the quasi-particle which comprises
all types correlations. Excited states, $\vert m; J^{\pi} M
\rangle$, in a compound nucleus are generated by operating the
following one phonon operator to the initial nucleus
\begin{equation} Q^{+,m}_{JM}  = {\mathop\Sigma_{k l \mu^{'}
\nu^{'} }} [ X^{m}_{(k \mu^{'} l \nu^{'} J)} C^{+}(k \mu^{'} l
\nu^{'} J M)
 - Y^{m}_{(k \mu^{'} l \nu^{'} J)} {\tilde C}(k \mu^{'} l \nu^{'}
  J M)] ~,\end{equation}
where pair creation operator $C^+$ is defined as
\begin{equation}
 C^{+} (k \mu^{'} l \nu^{'} J M)  =  {\mathop\Sigma_{m_{k} m_{l}}}
C^{JM}_{j_{k} m_{k} j_{l} m_{l}} a^{+}_{l \nu^{'}} a^{+}_{k
\mu^{'}}~,~ {\tilde C}(k \mu^{'} l \nu^{'} J M)  =  (-)^{J-M} C(k
\mu^{'} l \nu^{'} J - M)~
\end{equation}
with a quasi-particle creation operator $a^{+}_{l \nu^{'}}$ and
Clebsh-Gordan coefficient $C^{JM}_{j_{k} m_{k} j_{l} m_{l}}$. Here
Roman letters indicate single particle states, and Greek letters
with a prime mean quasi-particle types 1 or 2. If neutron-proton
pairing is neglected, the phonon operator is easily decoupled to
two phonon operators, charge changing and conserving reactions, as
used in the usual proton-neutron QRPA (pnQRPA) \cite{Ring08}. The amplitudes $X_{a
{\alpha}^{'}, b {\beta}^{'}}$ and $Y_{a {\alpha}^{'}, b
{\beta}^{'}}$, which stand for forward and backward going
amplitudes from ground states to excited states, are obtained from
the QRPA equation \cite{Ch09-2,Ch93}.

Under the second quantization, matrix elements of any transition
operator ${\cal {\hat O}}$ between a ground state and an excited
state $ | \omega ; J M >$ can be factored as follows
\begin{equation} < ~ \omega ; JM || {\cal {\hat O}}_{\lambda } || QRPA
>  =  {[\lambda]}^{-1} {\mathop{\Sigma}_{ab}} < a ||  {\cal {\hat O}}_{\lambda} || b>
<\omega ; J M  || {[c_a^+ {\tilde c}_b]}_{\lambda} ||  QRPA  > ~,
\end{equation}
where $c_a^+$ is the creation operator of a real particle at state
$a$. The first factor $< a ||{\cal {\hat O}}_{\lambda} || b
>$ can be calculated for a given single particle basis independently of the nuclear model
\cite{Don79,Wal75}. By using the phonon operator $Q^{+,m}_{JM}$,
we obtain the following expressions for CC and NC neutrino
reactions
\begin{eqnarray}
& &< ~ \omega ; JM  || {\cal {\hat O}}_{\lambda } ||  QRPA{>}_{CC}  \\
\nonumber = & & {\mathop\Sigma_{a \alpha^{'} b \beta^{'}}}  [
{\cal N}_{a \alpha^{'} b \beta^{'} } < a \alpha^{'} || {\cal {\hat
O}}_{\lambda}  || b \beta^{'}
>  ~[ u_{pa \alpha^{'}} v_{nb
\beta^{'}} X_{a \alpha^{'} b \beta^{'}} + v_{pa \alpha^{'}} u_{nb
\beta^{'}} Y_{a \alpha^{'} b \beta^{'}} ]~, \\ \nonumber
& & < ~ \omega ; JM  || {\cal {\hat O}}_{\lambda } || QRPA {>}_{NC}  \\
\nonumber = & & {\mathop\Sigma_{a \alpha^{'} b \beta^{'}}}  [
{\cal N}_{a \alpha^{'} b \beta^{'} } < a \alpha^{'} || {\cal {\hat
O}}_{\lambda}  || b \beta^{'}
>  ~[ u_{pa \alpha^{'}} v_{pb
\beta^{'}} X_{a \alpha^{'} b \beta^{'}} +
v_{pa \alpha^{'}} u_{pb \beta^{'}} Y_{a \alpha^{'} b \beta^{'}} ] \\
\nonumber & & - {(-)}^{j_a + j_b + J } {\cal N}_{ b \beta^{'} a
\alpha^{'} } <b \beta^{'}|| {\cal {\hat O}}_{\lambda}  || a
\alpha^{'}  >~ [ u_{pb \beta^{'}} v_{pa \alpha^{'}} X_{a
\alpha^{'} b \beta^{'}} +v_{pb \beta^{'}} u_{pa \alpha^{'}}  Y_{a
\alpha^{'} b \beta^{'}} ]] + (p \rightarrow n)~,
\end{eqnarray}
where $ {\cal N}_{a \alpha^{'} b \beta^{'}} (J) =  {\sqrt{ 1 +
\delta_{ab} \delta_{\alpha^{'}  \beta^{'} } (-1)^{J} }}/ ({1 +
\delta_{ab}\delta_{\alpha^{'}  \beta^{'} } }) $. By switching off
the np pairing, these forms are also easily reduced to the result
by the pnQRPA \cite{Ring08}

The weak current operators, ${\hat O}_{\lambda }$, composed of
longitudinal, Coulomb, electric and magnetic operators, are
detailed at Ref. \cite{Ch09-2}. Finally, based on initial and
final nuclear states, cross section for $\nu ({\bar \nu})$
reactions through the weak transition operator is given as
\cite{Wal75}
\Be & & ({{d \sigma_{\nu}} \over {d \Omega }  })_{(\nu / {\bar
\nu})} = { { G_F^2 \epsilon k } \over {\pi ~ (2 J_i + 1 ) }}~
\bigl[ ~ {\mathop\Sigma_{J = 0}} (
 1+ {\vec \nu} \cdot {\vec \beta }){| <  J_f || {\cal {\hat M}}_J || J_i > | }^2
 \\ \nonumber & & + (
 1 - {\vec \nu} \cdot {\vec \beta } + 2({\hat \nu} \cdot {\hat q} )
 ({\hat q} \cdot {\vec \beta}  ))
  {| <  J_f || {\cal {\hat L}}_J ||
J_i > | }^2  - \\ \nonumber & & {\hat q} \cdot ({\hat \nu}+ {\vec
\beta} )  { 2 Re < J_f || {\cal {\hat L}}_J  || J_i>
{< J_f|| {\cal {\hat M}}_J || J_i >}^*  } \\
\nonumber & &  + {\mathop\Sigma_{J = 1}} ( 1 - ({\hat \nu} \cdot
{\hat q} )({\hat q} \cdot {\vec \beta}  ) ) ( {| <  J_f || {\cal
{\hat T}}_J^{el}  || J_i > | }^2 + {| <  J_f || {\cal {\hat
T}}_J^{mag} || J_i > | }^2
) \\
\nonumber & &  \pm {\mathop\Sigma_{J = 1}} {\hat q} \cdot ({\hat
\nu} - {\vec \beta} )  2 Re [ <  J_f || {\cal {\hat T}}_J^{mag} ||
J_i > {<  J_f || {\cal {\hat T}}_J^{el} || J_i > }^* ]\bigr]~, \Ee
where $(\pm)$ means the case of $\nu ({\bar \nu})$, respectively.
${\vec \nu}$ and $ {\vec k}$ are incident and final lepton
3-momenta, ${\vec q} = {\vec k} - {\vec \nu}$ and ${\vec \beta} =
{\vec k} / \epsilon $ with the final lepton's energy $\epsilon$.

For CC reactions we multiplied Cabbibo angle $cos ^2 \theta_c$ and
considered the Coulomb distortion of outgoing leptons in a
residual nucleus \cite{Suzuki06,Ring08}. By following the
prescriptions on Refs.\cite{Ring08,Kolbe03-a}, we choose an energy
point in which both approaches predict same values. Then we use
the Fermi function below the energy and the effective momentum
approach (EMA) above the energy.

Our QRPA includes not only proton-proton and neutron-neutron
pairing but also neutron-proton (np) pairing correlations. The np
pairing is included at the BCS stage to reproduce empirical np
pairing gaps by adjusting a renormalized strength parameter
$g_{np}$ embedded in the Brueckner G matrix in the following
theoretical pairing gap \cite{Ch93}
\begin{equation}
\delta_{np} = - [ (H_0^{'} + E_1^{'} + E_2^{'}) - (H_0 + E_1 +
E_2) ],
\end{equation}
where $H_0^{'}(H_0)$ is total ground state energy with (without)
np pairing and $E_1^{'} + E_2^{'}~ ( E_1 + E_2)$ is a sum of the
lowest two quasi-particles energies with (without) np pairing
correlations. More detailed procedures for the inclusion of the np
pairing correlations are presented at Refs. \cite{Ch93,Ch10}.

\section{Results}
\vskip0.5cm
\subsection{Neutrino reactions on $^{40}$Ar via charged current}

In Fig.1, we show results for CC reactions, $^{40}$Ar($\nu_e,
e^-)$$^{40}$K$^{*}$ for solar and SN neutrinos. For solar
neutrinos (upper part), cross sections are fully ascribed to the
the GT and Fermi transitions. Other transitions contribute only a
few \% to the cross section below 30 MeV region. Since the maximum
energy of the solar neutrino from $^{8}$B is believed to be 12
MeV, solar neutrino reactions are dominated by the GT and Fermi
transitions.

These results justify previous SM calculations for solar neutrino
reactions on $^{40}$Ar \cite{Warb91,Orm95}, which consider only
the GT and Fermi transitions. The phenomenological approach
$\sigma =3 \sigma_{Fermi}$ used at Ref. \cite{Burrow} is
approximately good on the energy region above 20 MeV, but $\sigma
=2 \sigma_{Fermi}$ is much better below 20 MeV region.

But for SN neutrinos (lower part), contributions from the spin
dipole resonance (SDR) ($1^{-}$ and $2^{-}$) are increased to
those by the Fermi transition around 55 MeV and becomes larger
about 2 $\sim$ 2.5 times of the Fermi transition beyond the
region. Other higher multipole transitions, ($2^{\pm}, 3^{\pm}$
and $4^{\pm}$) contribute 10 \% maximally. Of course, both GT and
Fermi transitions are still main components. For example, their
contributions are 70 $\sim$ 80 \% around 50 MeV region, and 60
$\sim$ 50 \% above 50 MeV region. These phenomena, the dominance
of the GT and Fermi transitions, are typical of CC reactions on
even-even nuclei, for example, $^{12}$C, $^{56}$Fe and $^{56}$Ni
\cite{Ch09-1,Ch09-2}.

Our cross sections in Fig. 1 are smaller on the whole energy
region about 3.5 times rather than previous RPA calculations
\cite{Rub08}. In the following the characteristics of both
approaches are explained with the reason why we predict such small
cross sections. We include excited states of $^{40}$K$^{*}$ up to
a few tens of MeV, which are generated by the phonon operator in
Eq.(2), while the RPA calculation \cite{Rub08,Rub03} seems to take
only a few excited states known in the experiment.

But the high-lying excited states affect phase space in Eq. (6) through the energy conservation,
$\epsilon = E_{\nu} - \omega$, with the outgoing lepton energy $\epsilon$, the incident neutrino energy $E_{\nu}$
and the energy transfer $\omega$.
Energy transfer to the target nucleus $\omega$, which just corresponds to the excited energy,
makes the energy of outgoing electrons $\epsilon$ in Eq. (6) smaller than that of incident neutrino energy.
It leads to the smaller cross sections because the increase rate of transition strengths cannot follow the decrease of the phase space.


If we take the few known states, the cross sections are easily
increased to match well with the RPA calculation simply because of
the increase of the outgoing lepton energy in Eq.(6). Excited
states beyond the known states decrease cross sections even for
the solar neutrino case. Recent calculation of $^{40}$Ar($\nu ,
e^-$) by the local density approximation shows also such a
tendency because it does not also include discrete states
explicitly \cite{Athar06}.
\vskip0.5cm
\subsection{Gamow Teller strength by $^{40}$Ar (p,n) reactions}

In Fig. 2, we show the GT($\pm$) strength distributions and their
running sums, whose excitations energies are with respect to the
ground state of $^{40}$Ar. The higher energy states we go to, the
larger strength and their running sums are obtained. In specific,
discrete energy states on the 10 $\sim$ 20 MeV region affect
significantly the neutrino reaction. Even for solar neutrinos,
these energy states reduce cross sections compared to those
calculated by a few known states.

Very recently, the GT strength up to 8 MeV are extracted from
$^{40}$Ar(p,n) reactions \cite{Bhatt09}. Consistency of our
results with the experimental GT(--) strength can be shown by
studying the strength distribution and its running sum in the
lower energy states at Fig.2. The GT strength of ours are
localized at 3.6, 3.7, 8.6 and 8.9 MeV from the $^{40}$Ar ground
state, which corresponds to 2.1, 2.2, 7.1 and 7.4 MeV from the
$^{40}$K ground state. They are consistent with the data by the
$^{40}$Ar(p,n) reaction \cite{Bhatt09}, apart from their strength,
although some minor states between 2 and 7 MeV energy region are
not verified in detail in our calculations. But its running sum up
to 7.5 MeV from $^{40}$K ground state, whose value is $\Sigma ~
B(GT_{-}) = 4 \sim $5, is nicely reproduced in our approach by
multiplying the ${(g_V/g_A)}^2$ to our results $\Sigma ~ B(GT_{-})
=$ 3.6. The GT strength derived by our QRPA is thought to be
reliable enough to justify our conjecture on the role of higher
excited states.

Based on the GT and Fermi strength distributions, Bhattacharya
{\it et al.} \cite{Bhatt09} deduced $\nu_e$ reactions on $^{40}$Ar
up to 30 MeV in the following way
\begin{equation}
\sigma (E_\nu) = {{ G_F^2 cos^2 \theta_{c}} \over \pi \hbar^4 c^3}
{\mathop\Sigma_{i}}~ k_i \epsilon_i F(Z,\epsilon_i) [ B_i (GT) +
B_i (F)]~,
\end{equation}
where $k_i$ and $\epsilon_i$ refer to the momentum and total
energy of the outgoing electron and $F(Z,\epsilon_i)$ accounts for
the Coulomb correction. They are shown to be consistent with those
by the $\beta$ decay on the mirror nucleus, $^{40}$Ti
\cite{Bhatt98,Trinder97}. The neutrino cross sections of
$^{40}$Ar($\nu , e^-$) obtained by Eq.(8) were very similar to the
previous results calculated by a few experimental excited states
\cite{Rub08} around a few MeV region. For example, the cross
section around 30 MeV region is about $ 200 \times 10^{-42} cm^2$,
which is about 3 times larger than our prediction at Fig.1. Since
the excited states newly verified in the $^{40}$Ar(p,n) reaction
are also located on the energy region below 10 MeV, deduced cross
sections are rarely affected by the new GT strength data.

It should be noted that Eq. (8) is valid only near threshold of the outgoing electron as commented at Ref. \cite{Bhatt09}.
For example, one can obtain very easily Eq. (8) from Eq. (6), which is exact form for the cross section, by putting the 3-momentum of the outgoing electron ${\vec k}$ = 0 and considering only J = 0 and 1 for the Fermi and the GT transitions, respectively.
Therefore, Eq. (8) cannot be applied for the incident neutrino energy beyond the electron threshold considered in this report.

\vskip0.5cm
\subsection{Anti-neutrino reactions on $^{40}$Ar via charged current}

This reduction of cross sections by the higher excited states
should function on the ${\bar \nu}_e$ reaction, $^{40}$Ar(${\bar
\nu}_e, e^+)$$^{40}$Cl$^{*}$. Fig. 3 shows results for
$^{40}$Ar(${\bar \nu}_e, e^+)$$^{40}$Cl$^{*}$ via CC for solar and
SN neutrinos. General trends of the GT, Fermi and SDR transitions
are very similar to the results of $^{40}$Ar(${\nu}_e,
e^-)$$^{40}$K$^{*}$. For solar neutrinos, the GT and Fermi
transitions are main components, but the SDR ($1^-$ and $2^-$)
contributions emerge largely for the SNe neutrinos. Cross sections
for solar ${\bar \nu}_e$ are almost same as the those by the
$\nu_e$ reaction, $^{40}$Ar(${\nu}_e, e^-)$$^{40}$K$^{*}$, while
they are about a half of those by the $\nu_e$ reaction for SN
neutrinos.

This is contrast to the results of Refs. \cite{Rub08,Rub03}.
Actually, they addressed about 12 times difference between the
$\nu_e$ and ${\bar \nu}_e$ reactions in the cross section at
$E_{\nu}$ = 80 MeV. Since the Q value for the ${\bar \nu}_e$
reaction, 7.48 MeV, decreases the incident neutrino energy by the
Q value, {\it i.e.} as $E^i_{\nu} \rightarrow E^i_{\nu} - Q = \epsilon + \omega$, the
reaction is disfavored by the low energy neutrino. Consequently,
it may play a role of reducing the ${\bar \nu}_e$ cross section
for a given energy. The strong reduction of cross sections for the
${\bar \nu}_e$ reaction by the large Q value was one of the
motivation to distinguish the $\nu_e$ and ${\bar \nu}_e$ reactions
on $^{40}$Ar target.

However, the high-lying excited states weaken the decrease of cross
sections by the Q value, so that the expected large decrease of
cross sections by the Q value is not so drastic compared to those
by a few known excited states. As a result, our cross sections by
${\bar \nu}_e$ are about twice larger than those of Refs.
\cite{Rub03,Rub08}. Since the cross sections by $\nu_e$ become
smaller about 3.5 times than those by previous calculations, total
difference between the $\nu_e$ and ${\bar \nu}_e$ reactions turns
out to be only 2 $\sim$ 3 times difference. Higher excited states
are directly associated to the reason why we have only about twice
difference at $E_{\nu}$ = 80 MeV between the $\nu_e$ and ${\bar
\nu}_e$ reactions.

In general, ${\bar \nu}_e$ cross sections up to around a few tens
MeV region are nearly same as those by ${\nu}_e$, if we
investigate results for other nuclei. Actually, the difference
between ${\bar \nu}_e$-A and ${\nu}_e$-A reactions is given by the
last term in Eq.(6), the interference term of magnetic and
electric transitions. It means that main interactions for
$\nu$--$^{40}$Ar stem from the longitudinal and Coulomb
transitions. Nuclear effects such as the Q value are not so large
as to give rise to such a large difference, if we consider the
higher excited states.


Fermi function for the Coulomb correction is used on the energy
region below 40 MeV and the effective momentum approach (EMA) is
taken beyond 40 MeV. Coulomb corrections do not affect the
difference between ${\bar \nu}_e$-A and ${\nu}_e$-A reactions.
Both reactions are increased by about 15 \% maximally by the
Coulomb distortion. For pairing interactions, $g_{nn} = 1.105,
g_{pp} = 1.057$ and $g_{np} = 1.5368$ are used to reproduce the
empirical pairing gaps $\Delta_{nn} =1.768, \Delta_{pp} = 1.776$
and $\delta_{np} =0.684$ MeV for $^{40}$Ar, respectively
\cite{Ch93}.

\vskip0.5cm
\subsection{Neutrino and anti-neutrino reactions on $^{40}$Ar via neutral current}

Figs.4 and 5 shows results for NC reactions $^{40}$Ar($\nu_e,
\nu_e)$$^{40}$Ar$^{*}$ and $^{40}$Ar(${\bar \nu}_e, {\bar
\nu}_e)$$^{40}$Ar$^{*}$ for solar and SN neutrinos. They are
dominated by the GT transition for solar neutrinos, which accounts
for about a half of cross sections for SN neutrinos. It is typical
of the NC reaction on even-even nuclei \cite{Ch09-1,Ch09-2}. Cross
sections presented here are smaller than those used in Refs.
\cite{Rub03,Rub08} because of the roles of higher energy states
discussed for the CC reactions. One more point to be noticed is
that our calculations for CC and NC reactions are carried out by a
framework as shown in Eqs. (2) and (5). All of results are
summarized in Fig.6 by the logarithm to grasp whole neutrino
reactions on $^{40}$Ar. Cross sections by the CC are 4 $\sim$ 5
times larger than those by the NC reaction

\section{Summaries and Conclusion}
We calculated neutrino induced reactions on $^{40}$Ar by including
multipole transitions up to $J^{\pi} = 4^{\pm}$ with explicit
momentum dependence. Our QRPA includes neutron-proton (np) pairing
as well as neutron-neutron and proton-proton pairing correlations.
Since energy gaps between proton and neutron energy spaces in
medium nuclei are adjacent to each other, the np pairing may
affect significantly the nuclear weak interaction.

In this work, we take account of excited states up to a few tens
of MeV in contrast to the previous RPA calculations, which
consider only a few known states in the experiment. These excited
states, in specific, located on 10 $\sim$ 20 MeV region reduce
strongly corresponding cross sections and weaken the decrease of
the cross sections by the Q value in the ${\bar \nu}_e$ reaction.

The amount of the increase of cross sections in our results is
smaller than that by previous calculations. They consider only
low-lying states and maybe treat the neglected high-lying states
as elastic channels, which leads to such large cross sections.
Contributions by the elastic channels in high-lying states in
previous calculations, which are usually larger than inelastic
channels, for example, the quasi-elastic peak, are compensated by
the explicit contribution owing to the inelastic channels coming
from the high-lying states.

By these higher excited states, the large Q value for
$^{40}$Ar(${\bar \nu}_e, e^+)$$^{40}$Cl$^{*}$ reaction, which
motivated discerning $\nu_e$ and ${\bar \nu}_e$ reactions on
$^{40}$Ar, does not give rise to such a drastic difference about
12 times, but it leads to about 2 times difference between both
reactions. More experimental data for higher excited states
through relevant experiments are necessary for further discussion
and conclusion for the difference. Study of the GT transition by
($^{3}$He, t) or (p,n) reactions could be viable approaches for
obtaining the data \cite{OMEG10}.

$^{40}$Ar has a difficulty of considering two different major
shells, $sd$ and $fp$, which cause large number configurations
with multi-particle and multi-hole interactions. The QRPA is a
very efficient method to consider such multi-particle and
multi-hole interactions and their configuration mixing, and
successfully described the nuclear weak reactions sensitive on the
nuclear structure, such as various $\beta$ decays and relevant
neutrino-nucleus reactions. Therefore our QRPA results for the
$\nu_e ({\bar \nu}_e) - ^{40}$Ar reaction could be a useful
reference for the detection of SN neutrinos in the LArTPC
detector.

Possible deformation in the exotic nuclei of astrophysical
importance might be dealt with the Deformed QRPA (DQRPA)
\cite{saleh}, which explicitly takes the deformation into account
in the deformed Nilsson basis. Preliminary results on the GT
strength distribution by the DQRPA show the importance of the
deformation for understanding the neutrino induced reactions on
the unstable nuclei \cite{Ha10}.

This work was supported by the National Research Foundation of
Korea (2011-0015467) and one of author, Cheoun, was supported by
the Soongsil University Research Fund.


\begin{figure}
\includegraphics[width=0.85\linewidth]{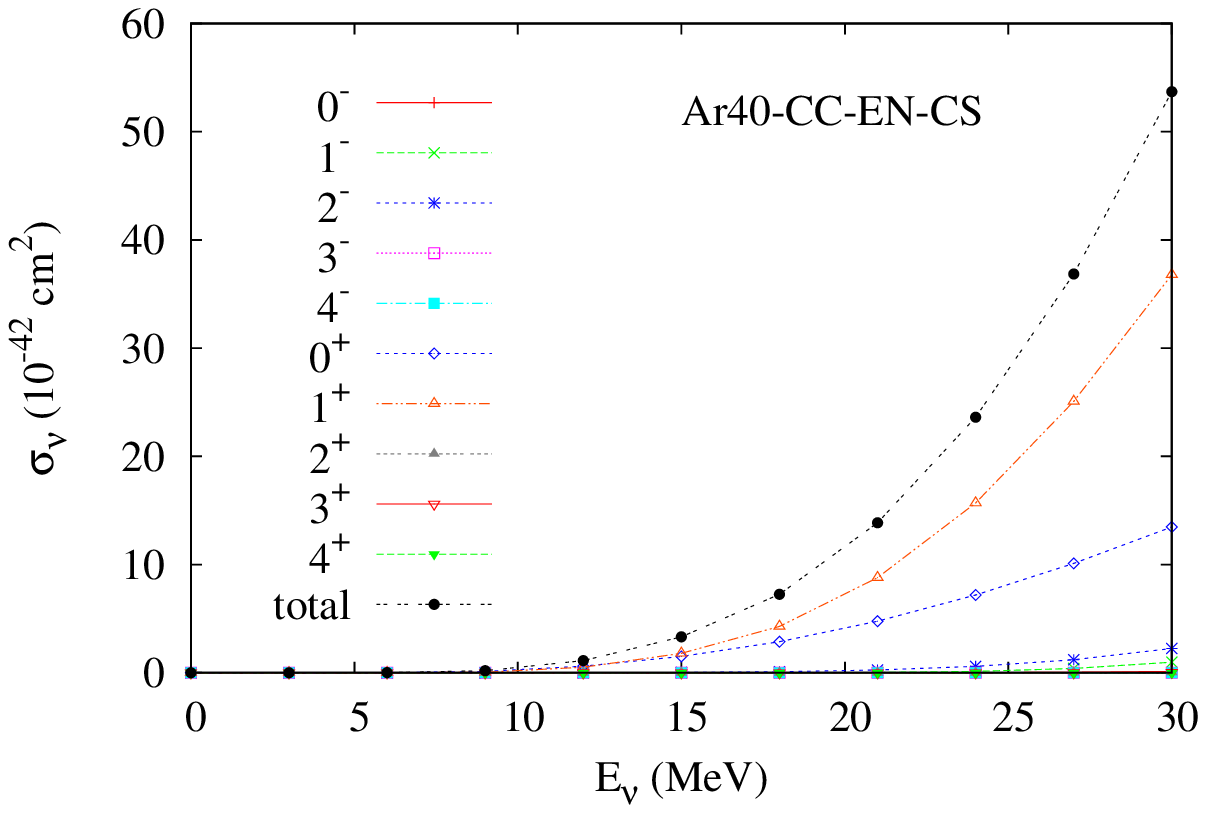}
\includegraphics[width=0.85\linewidth]{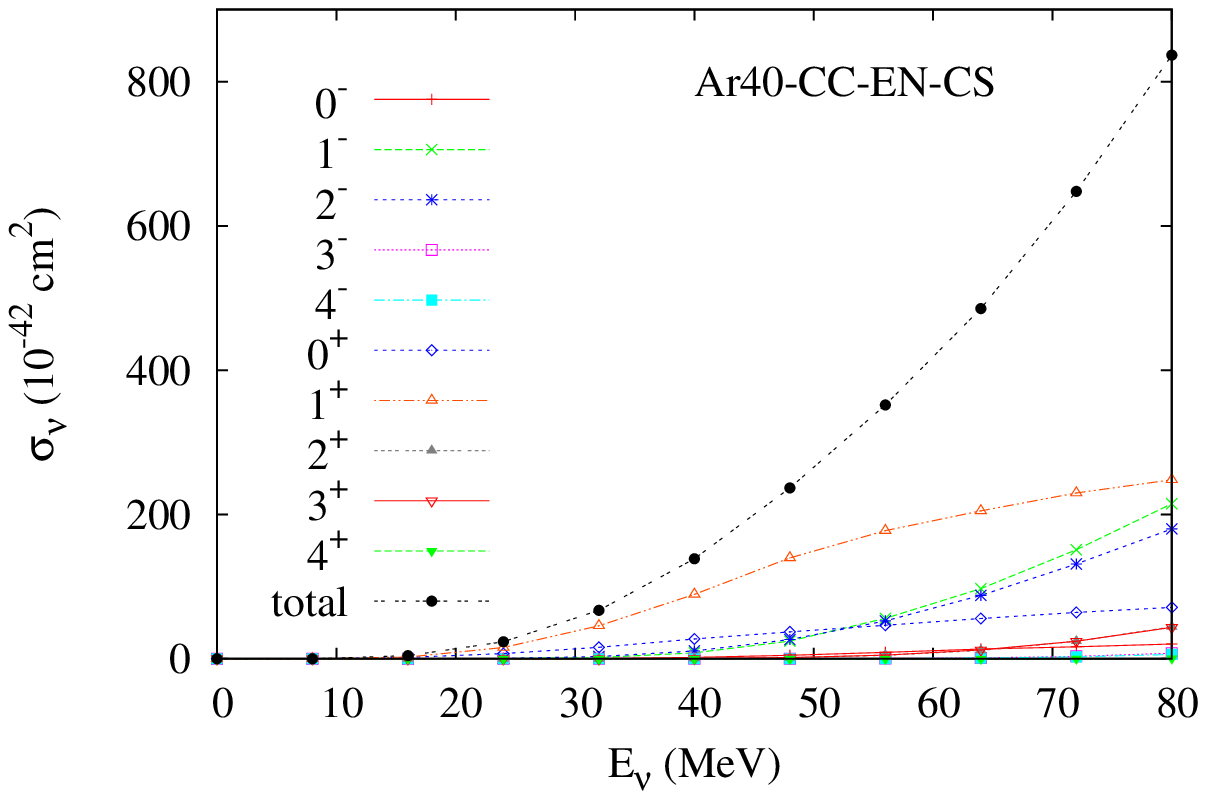}
\caption{(Color online) Cross sections by charged current,
$^{40}$Ar($\nu_e, e^-)$$^{40}$K$^{*}$ for solar (upper) and SN
(lower) neutrinos. For solar (supernova) neutrinos, the
$E_{\nu_e}^{max.} = 30 (80) $ MeV are used for $J_{\pi} = 0^{\pm}
\sim 4^{\pm}$ states.}
\label{fig1}
\end{figure}


\begin{figure}
\centering
\includegraphics[width=7.5cm]{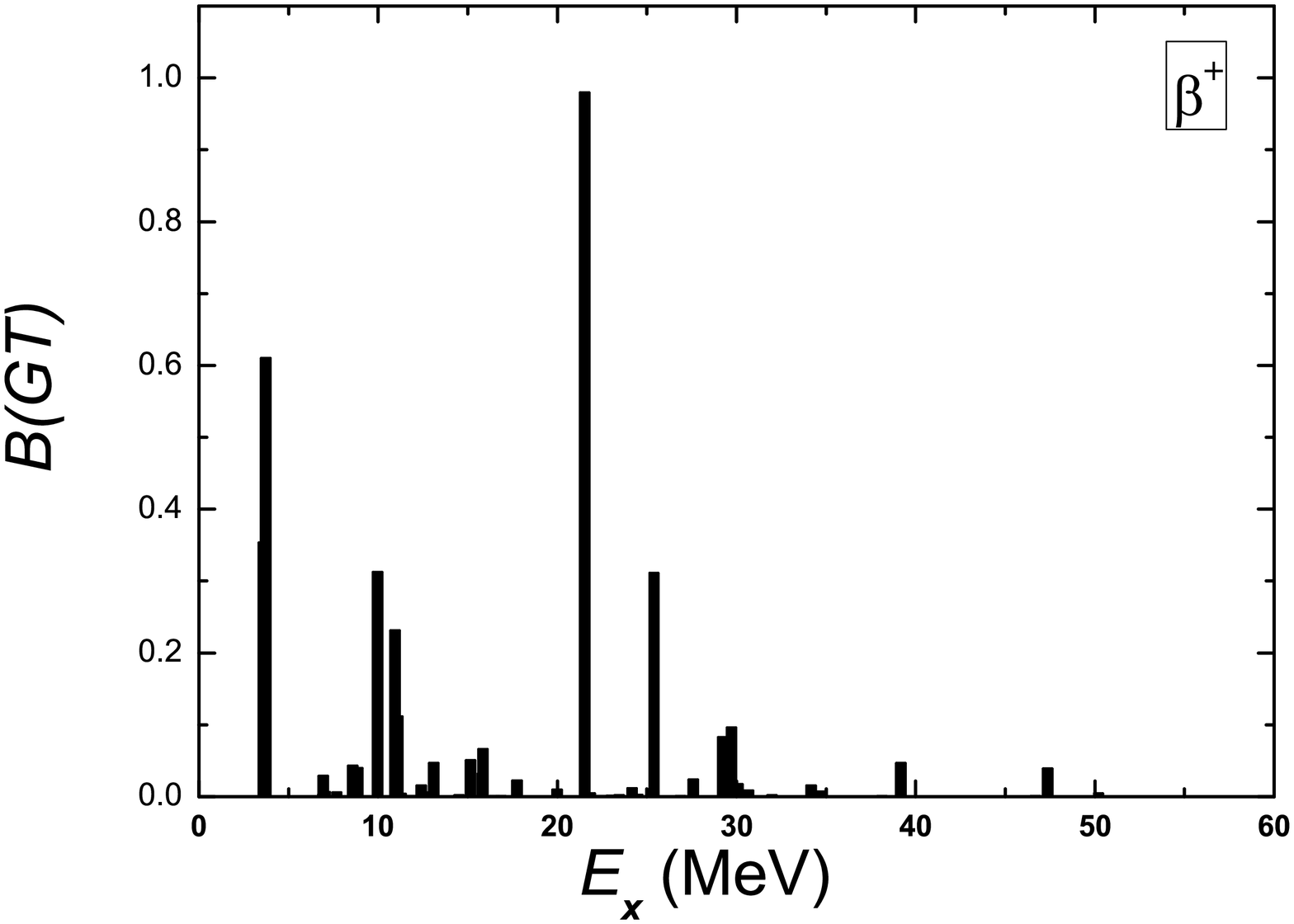}
\includegraphics[width=7.5cm]{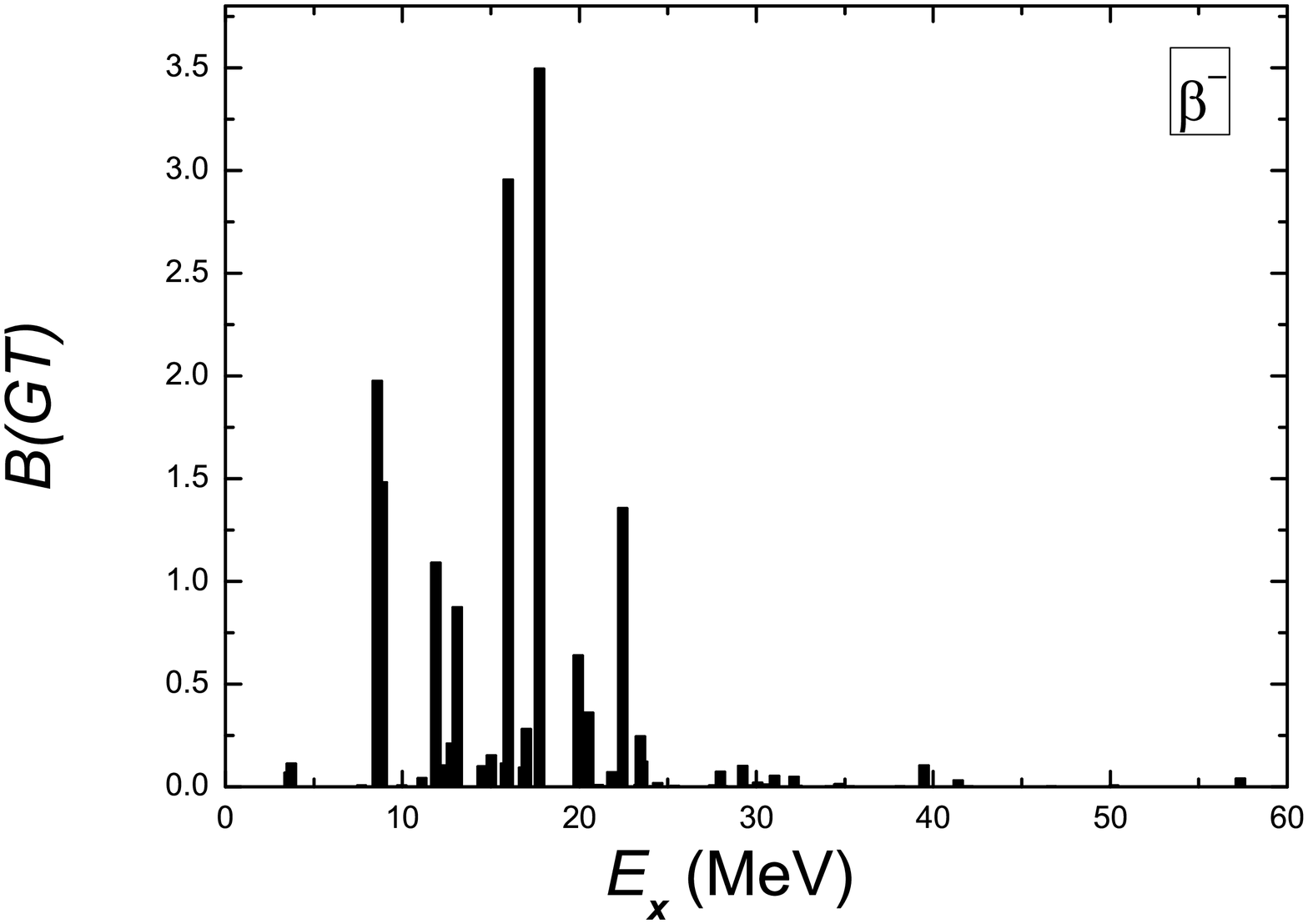}
\includegraphics[width=7.5cm]{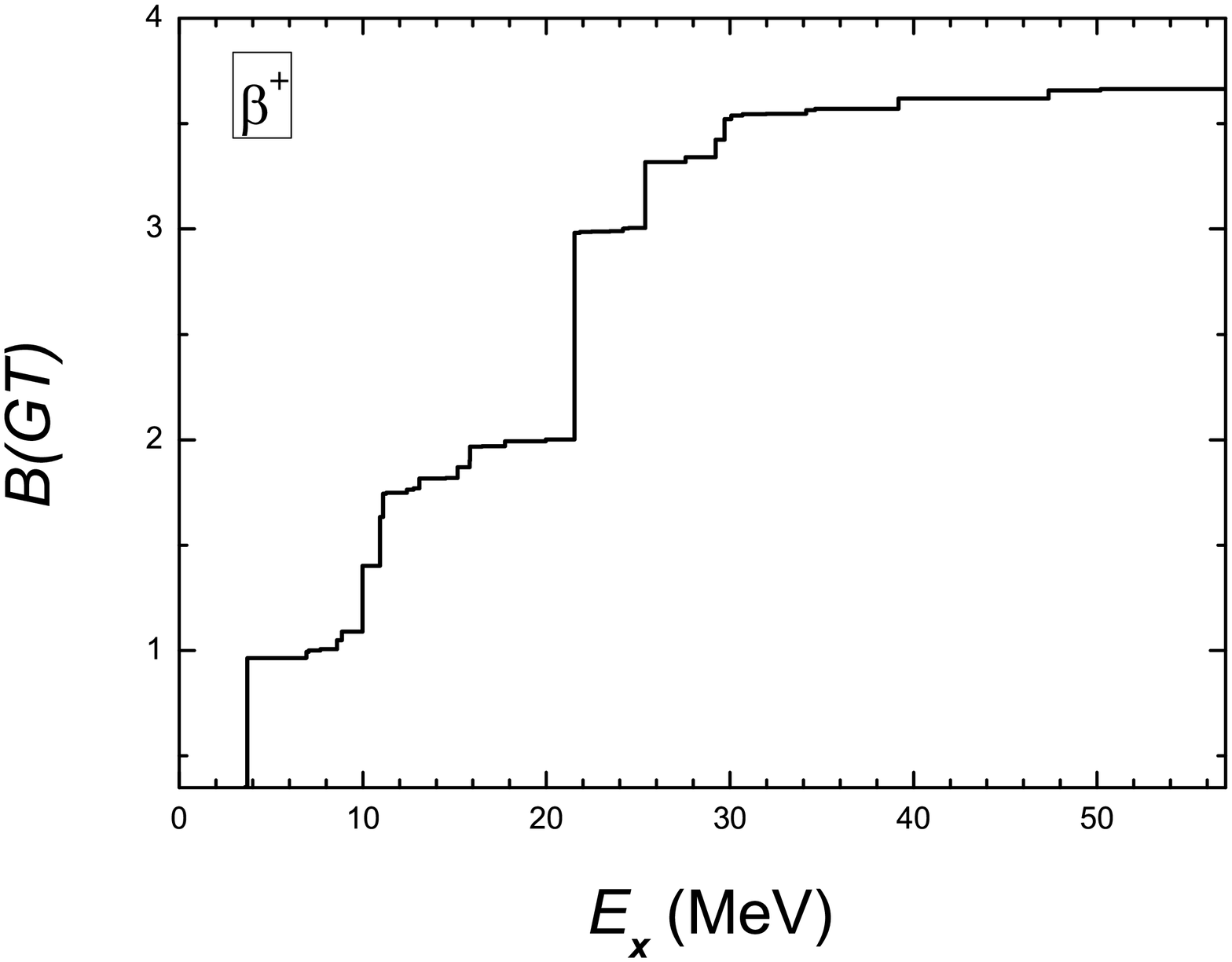}
\includegraphics[width=7.5cm]{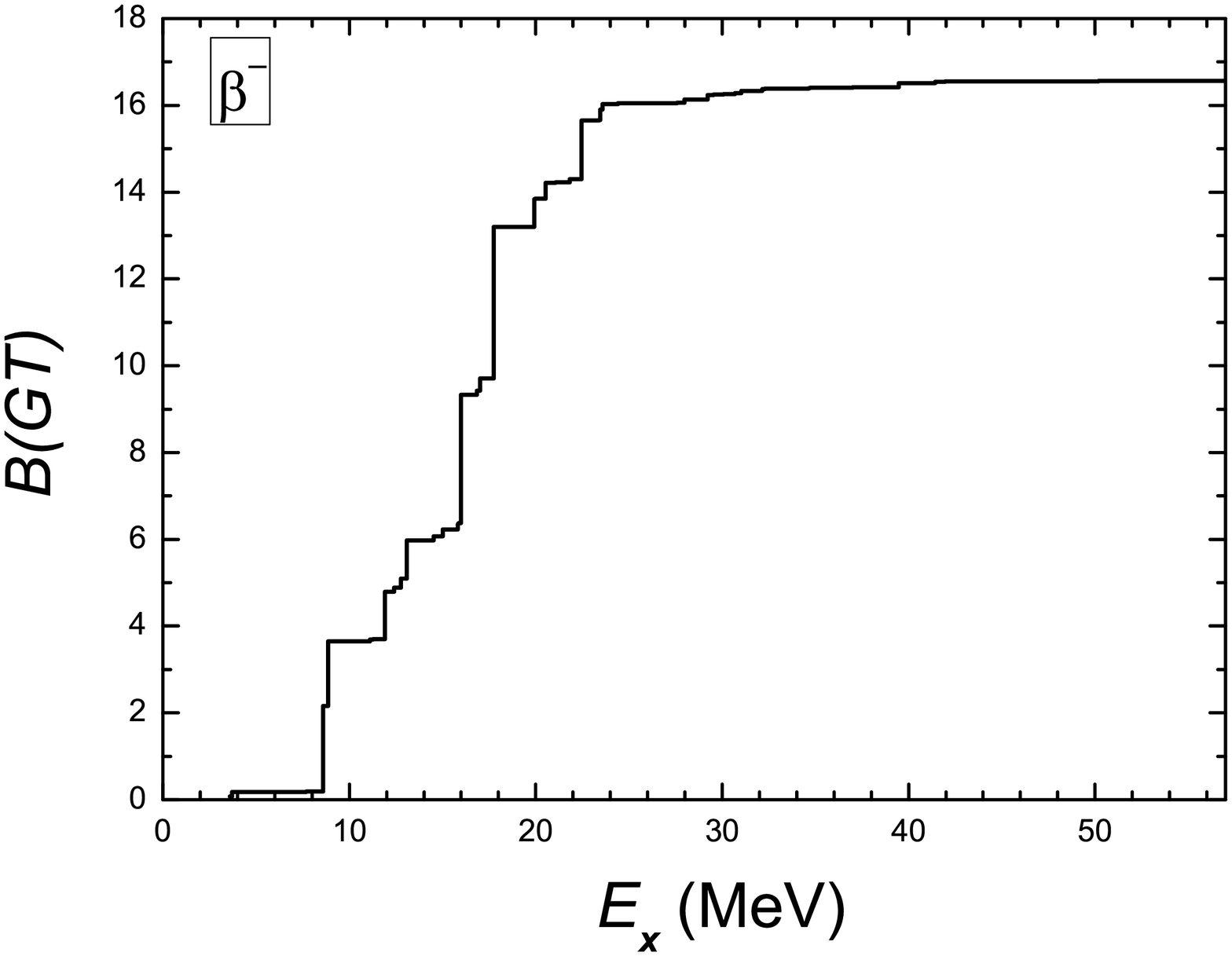}
\caption{ The Gamow Teller strength GT$({\pm})$ from $^{40}$Ar and
their running sums. $E_x$ is with respect to the ground state of
$^{40}$Ar.}\label{fig2}
\end{figure}

\begin{figure}
\includegraphics[width=0.85\linewidth]{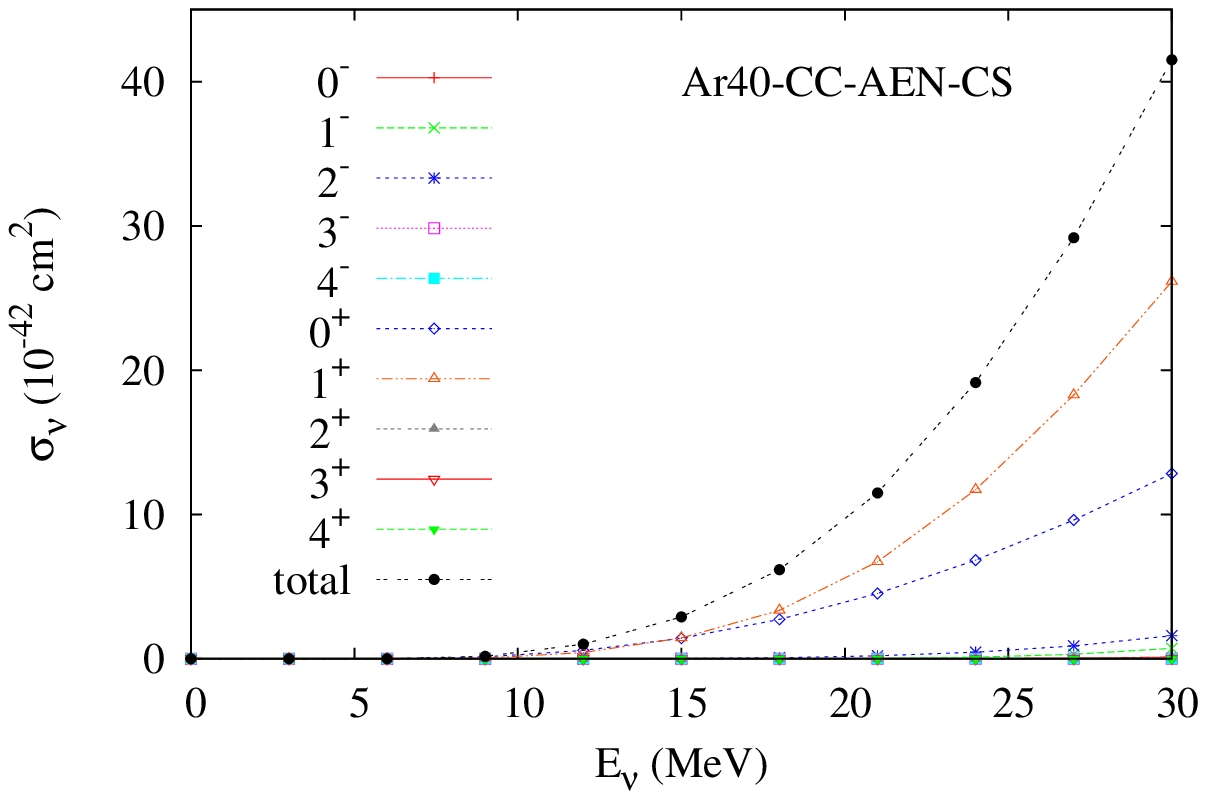}
\includegraphics[width=0.85\linewidth]{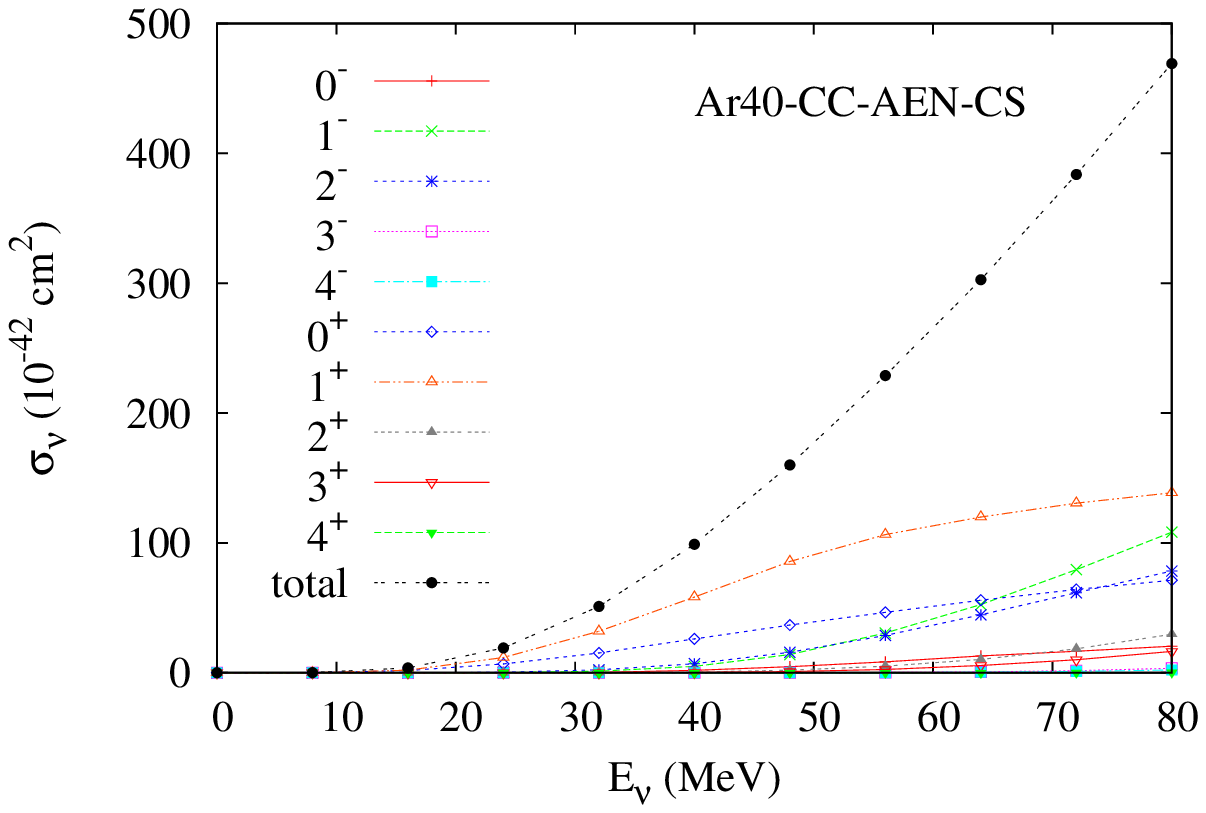}
\caption{(Color online) Cross sections by charged current,
$^{40}$Ar(${\bar \nu}_e, e^+)$$^{40}$Cl$^{*}$ for solar (upper)
and SN (lower) neutrinos.} \label{fig3}
\end{figure}

\begin{figure}
\centering
\includegraphics[width=0.85\linewidth]{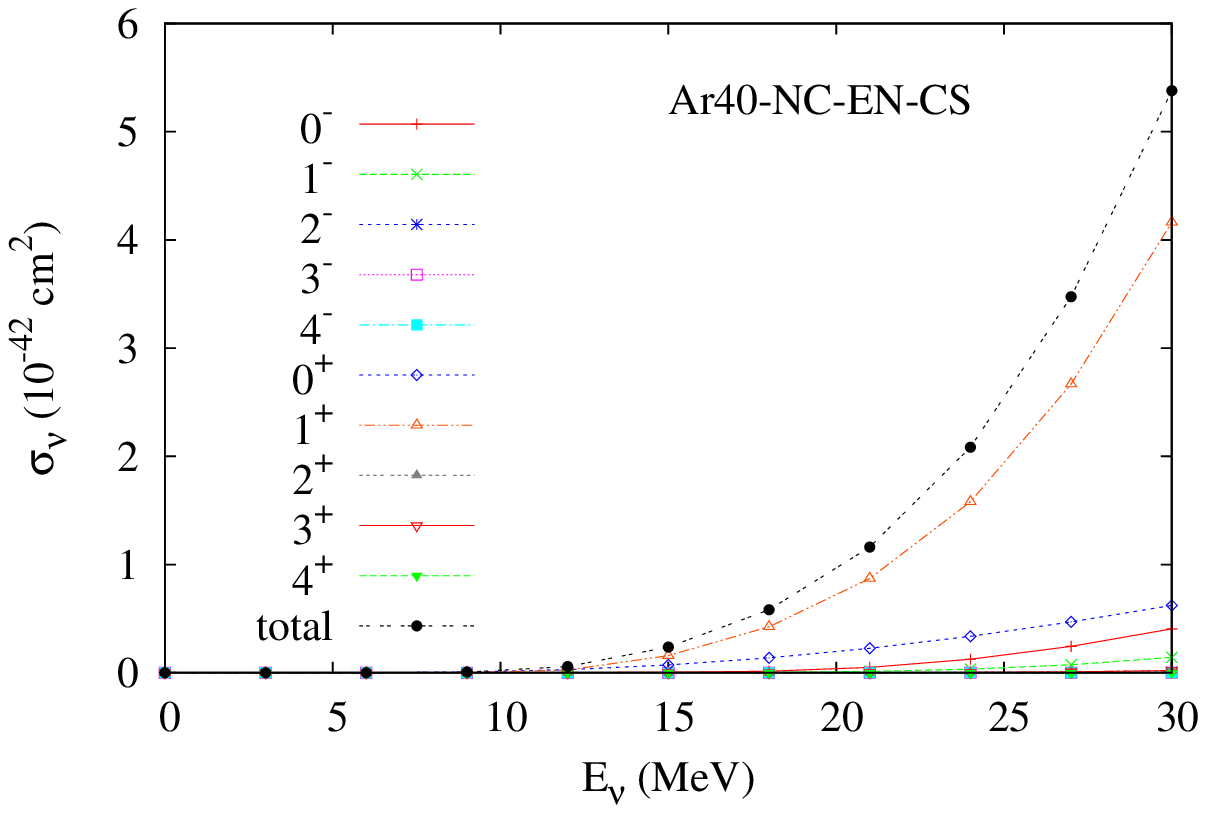}
\includegraphics[width=0.85\linewidth]{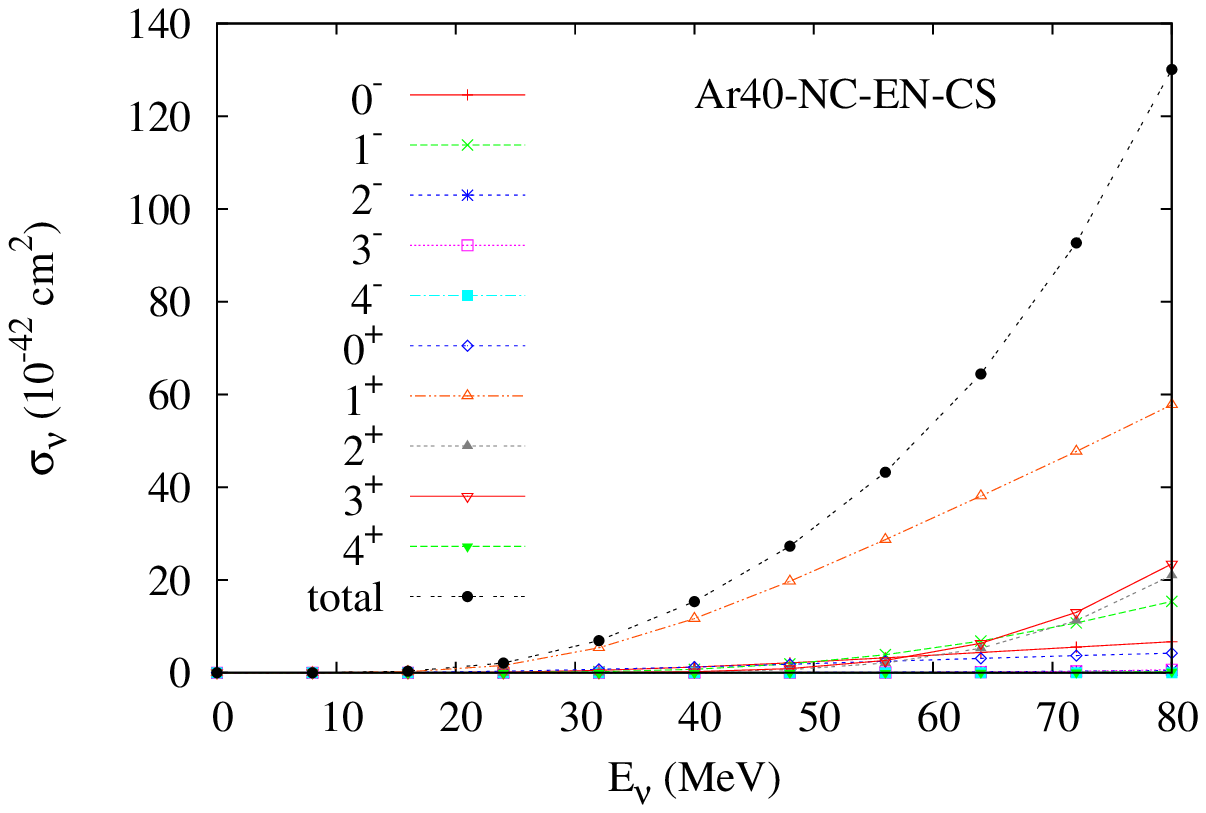}
\caption{(Color online) Cross sections by neutral current,
$^{40}$Ar($\nu_e, \nu_e^{'})$$^{40}$Ar$^{*}$ for solar (upper) and
SN (lower) neutrinos.}\label{fig4}
\end{figure}

\begin{figure}
\centering
\includegraphics[width=0.85\linewidth]{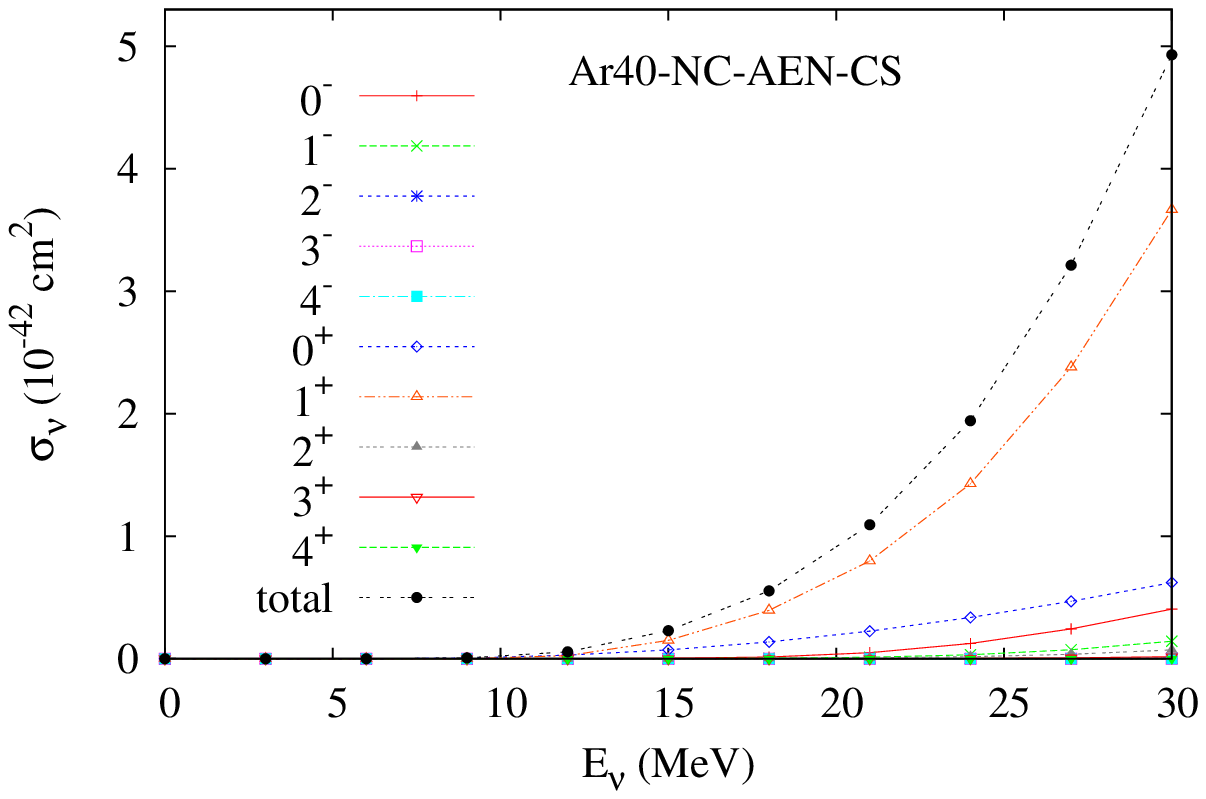}
\includegraphics[width=0.85\linewidth]{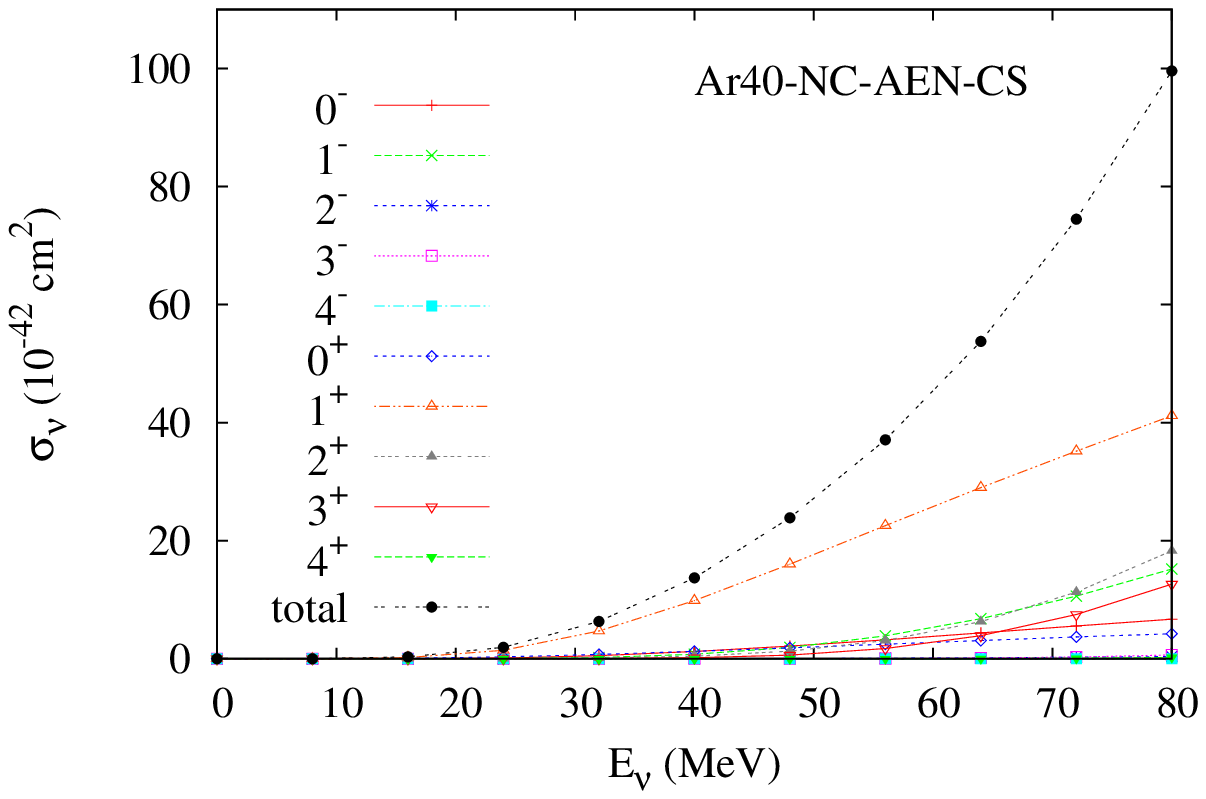}
\caption{(Color online) Cross sections by neutral current,
$^{40}$Ar($\nu_e, \nu_e^{'})$$^{40}$Ar$^{*}$ for solar (upper) and
SN (lower) neutrinos.}\label{fig5}
\end{figure}

\begin{figure}
\centering
\includegraphics[width=0.85\linewidth]{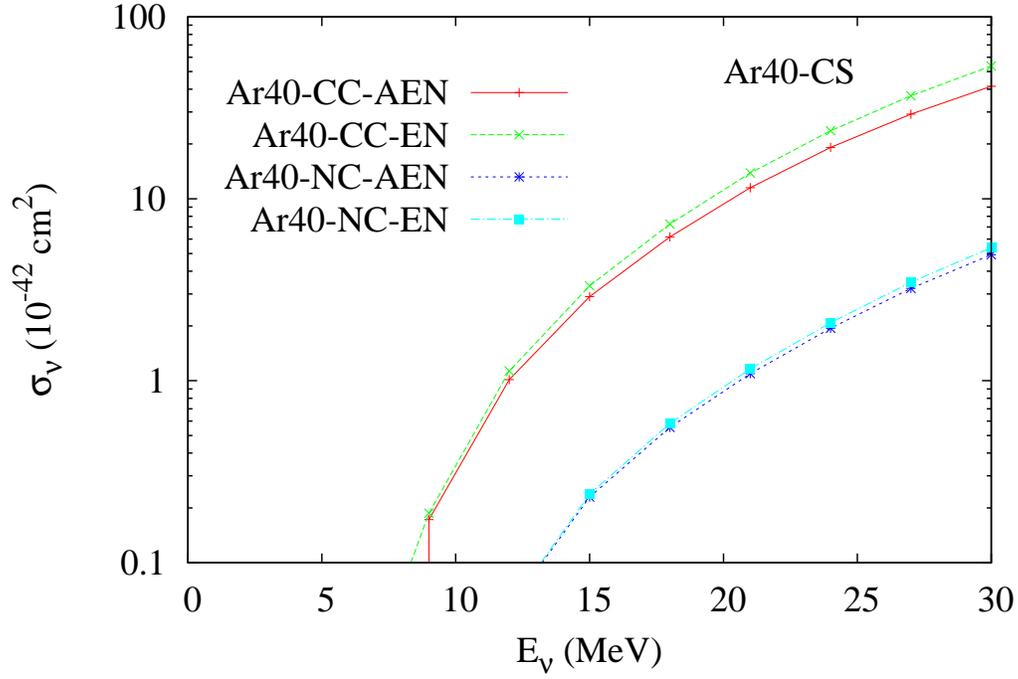}
\includegraphics[width=0.85\linewidth]{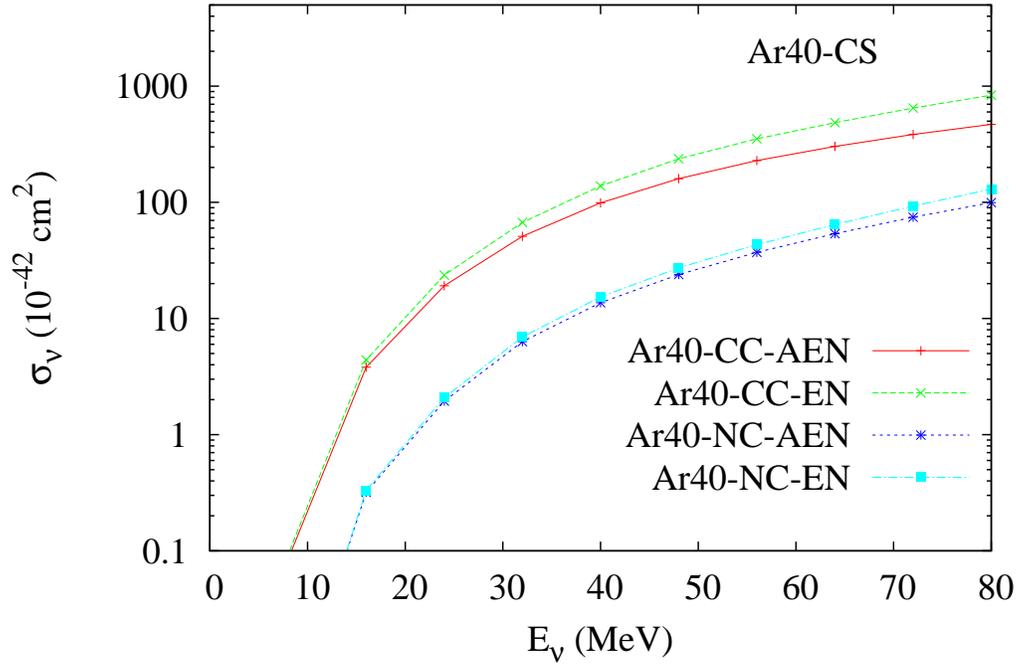}
\caption{(Color online) Comparison of cross sections for relevant
neutrino reactions on $^{40}$Ar given by the log scale for solar
(upper) and SN (lower) neutrinos.}\label{fig6}
\end{figure}


\begin{thebibliography}{140}
\par
\vskip0.5cm
\par
\def\pr{Phys. Rev.}
\def\prl{Phys. Rev. Lett.}
\def\nc{Nucl. Phys.}
\def\pl{Phys. Lett.}
\def\nuc{Nuovo. Cim.}
\def\pro{Prog. Theo. Phys}
\def\so{Sov.J.Nucl.Phys.}
\def\can{Can. J. Phys.}
\par

\bibitem{Rub98} A. Rubbia, Nucl. Phys. {\bf B 66}, 436 (1998).

\bibitem{Rub08} I. Gil-Botella and A. Rubbia, arXiv:hep-ph/0307244v2, Revised Feb.7, (2008).

\bibitem{Rub03} I. Gil-Botella and A. Rubbia, J. of Cosmology and
Astroparticle Physics, {\bf 10}, 009 (2003).



\bibitem{Woosley90} S. E. Woosley, D. H. Hartmann, R. D. Hoffmann,
and W. C. Haxton, Astrophys. J. {\bf 356}, 272 (1990).
\bibitem{Yoshida08} T. Yoshida, T. Suzuki, S, Chiba, T. Kajino, H.
Yokomukura, K. Kimura, A. Takamura, H. Hartmann, Astro. Phys. J.
{\bf 686}, 448 (2008).
\bibitem{Heg} A. Heger E. Kolbe, W.C. Haxton, K. Langanke, G. Mart{\'i}nez-Pinedo,
S.E. Woosley, Phys. Lett. {\bf B606}, 258 (2005).


\bibitem{Suzuki09} T. Suzuki, M. Honma, K. Higashiyama, T. Yoshida, T. Kajino, T.
Otsuka, H. Umeda, and K. Nomoto, Phys. Rev. C {\bf 79}, 061603(R)
(2009).


\bibitem{Suzuki06} T. Suzuki, S. Chiba, T. Yoshida, T. Kajino, T.
Otsuka, Phys. Rev. C {\bf 74}, 034307 (2006).

\bibitem{Ch11} Myung-Ki Cheoun, T. Kajino, M. Kusagabe, Grant J. Mathews, Phys. Rev. D {\bf 84}, 043001 (2011).
\bibitem{Smir00} A. S. Dighe and A. Y. Smirnov, Phys. Rev. {\bf D
62}, 033007 (2000).
\bibitem{Curi06} A. Curioni, Nucl. Phys. {\bf B 159}, 69 (2006).


\bibitem{Kolbe03-a} E. Kolbe, K. Langanke, G Martinez-Pinedo and P. Vogel, J. Phys.
G {\bf 29}, 2569 (2003).






\bibitem{Ch09-1} Myung-Ki Cheoun, Eunja Ha, S. Y. Lee, W. So, K. S. Kim and T.
Kajino, Phys. Rev. {\bf C 81}, 028501 (2010).
\bibitem{Ch09-2} Myung-Ki Cheoun, Eunja Ha, K. S. Kim and T.
Kajino, J. of Phys. {\bf G 37}, 055101, (2010).
\bibitem{Ch10} Myung-Ki Cheoun, Eunja Ha, T. Hayakawa, S. Chiba and T.
Kajino, Phys. Rev. {\bf C 82}, 035504, (2010).

\bibitem{Burrow} Todd A. Thompson, Adam Burrows, Philip A.
Pinto, Astrophys. J. {\bf 592}, 434 (2003).

\bibitem{Ch09-3} K. S. Kim and Myung-Ki Cheoun, Phys. Lett. {\bf B 679}, 330 (2009).
\bibitem{Ch09-4} Myung-Ki Cheoun and K. S. Kim, J. of the Phys. Soc. of Japan {\bf 78}, 084202 (2009).
\bibitem{Haus52} W. Hauser and H. Feshbach, Phys. Rev. {\bf 87}, 366
(1952).
\bibitem{Naka05} T. Nakagawa, S. Chiba, T. Hayakawa, T. Kajino,
Atomic Data and Nuclear Data Table, {\bf 91}, 77, (2005).
\bibitem{Ring08} N. Paar, D. Vretenar, T. Marketin, and P. Ring,
Phys. Rev. {\bf C 77}, 024608 (2008).
\bibitem{Ch11-2}  Myung-Ki Cheoun, Eunja Ha and T.
Kajino, Phys. Rev. {\bf C 83}, 028801, (2011).
\bibitem{Bhatt09} M. Bhattacharya, C. D. Goodman, and A. Garcia,
Phys. Rev. {\bf C 80}, 055501 (2009).




%



\bibitem{Ch93} M. K. Cheoun, A. Bobyk, Amand Faessler, F. Simcovic and
G. Teneva, {\nc} {\bf {A561}}, 74 (1993) ; {\nc} {\bf {A564}}, 329
(1993); M. K. Cheoun, G. Teneva and Amand Faessler, Prog. Part.
Nuc. Phys. {\bf 32}, 315 (1994) ; M. K. Cheoun, G. Teneva and
Amand Faessler, {\nc} {\bf A587}, 301 (1995).


\bibitem{Don79} T. W. Donnelly and W. C. Haxton, ATOMIC DATA AND
NUCLEAR DATA {\bf 23}, 103 (1979).

\bibitem{Wal75} J. D. Walecka, {\it Muon Physics}, edited by V.
H. Huges and C. S. Wu (Academic, New York, 1975), Vol II.


\bibitem{Warb91} E. K. Warburton, Phys. Rev. {\bf C 44}, 268
(1991).
\bibitem{Athar06} M. S. Athar, S. Ahmad, S. K. Singh, Nucl. Phys.
{\bf A 764}, 551 (2006).

\bibitem{Trinder97} W. Trinder {\it et al.}, Phys. Lett. {\bf B
415}, 211 (1997).
\bibitem{Bhatt98} M. Bhattacharya {\it et al.}, Phys. Rev. {\bf C
58}, 3677 (1998).


\bibitem{Orm95} W. E. Ormand, P. M. Puzzochero, P. F. Bortignon,
R. A. Broglia, Phys. Lett. {\bf 345}, 343 (1995).







\bibitem{OMEG10} Y. Shimbara {\it et al.}, {\it The 10th International Symposium on
Origin of Matter and Evolution of Galaxies}, edited by I. Tanihara
{\it et. al.}, AIP, New York, 201 (2010).

\bibitem{saleh} M. S. Yousef, V. Rodin, A. Faessler, and F. Simkovic,
Phys. Rev. C {\bf 79}, 014314 (2009).

\bibitem{Ha10} Eunja Ha and Myung-Ki Cheoun, {\it The 10th International Symposium on
Origin of Matter and Evolution of Galaxies}, edited by I. Tanihara
{\it et. al.}, AIP, New York, 351 (2010).
\end{thebibliography}
\end{document}